\documentclass[pra,aps,twocolumn]{revtex4-1}
\usepackage{graphicx} 
\usepackage{epsfig} 

\begin{document}

\title{Low-energy monopole strength in exotic Nickel isotopes}

\author{E. Khan}
\affiliation{Institut de Physique Nucl\'eaire, Universit\'e Paris-Sud, IN2P3-CNRS, F-91406 Orsay Cedex, France}
\author{N. Paar}
\author{D. Vretenar}
\affiliation{Physics Department, Faculty of Science, University of Zagreb, Croatia}

\begin{abstract}
Low-energy strength is predicted for the isoscalar monopole response of
neutron-rich Ni isotopes, in calculations performed using the
microscopic Skyrme HF+RPA and relativistic RHB+RQRPA models.  Both
models, although based on different energy density functionals,
predict the occurrence of pronounced monopole states in the energy
region between 10 MeV and 15 MeV, well separated from the isoscalar
GMR. The analysis of transition densities and corresponding
particle-hole configurations shows that these states represent almost
pure neutron single hole-particle excitations. Even though their
location is not modified with respect to the corresponding unperturbed
states, their (Q)RPA strength is considerably enhanced by the 
residual interaction. The theoretical analysis predicts the gradual
enhancement of low-energy monopole strength with neutron excess. 
\end{abstract}


\date{\today}

\maketitle

The multipole response of nuclei far from the $\beta$-stability line
and the possible occurrence of exotic modes of excitation presents a
growing field of research. Besides being intrinsically interesting as
new structure phenomena, exotic modes of excitation might play an
important role in r-process nucleosynthesis \cite{gor04}.  Low-lying
E1 strength has been measured in neutron-rich Oxygen, Neon, and Tin
isotopes \cite{gib08,lei01,adr05}, and more recently in $^{68}$Ni
\cite{wie09}. The interpretation of the dynamics of the observed
low-energy E1 strength in nuclei with a pronounced neutron excess is
currently under discussion (see Ref. \cite{paa07} for a recent
review). In light nuclei such as the Oxygen isotopes the onset of dipole
strength in the low-energy region is associated with non-resonant
independent single-particle excitations of  loosely bound neutrons.
However, in the case of $^{68}$Ni the low-lying dipole strength is
found to be rather collective \cite{vre01}, and several theoretical analyses
have predicted the existence of the pygmy dipole resonance (PDR) in
medium-mass and heavy nuclei, i.e., the resonant oscillation of the
weakly-bound neutron skin against the isospin saturated proton-neutron
core. More recently it has been shown that the low-lying E1 strength
can exhibit a non trivial pattern, separated into two segments with
different isospin character. The more pronounced pygmy structure at
lower energy is composed of predominantly isoscalar states with
surface-peaked transition densities. At somewhat higher energy the
calculated E1 strength is primarily of isovector character, as
expected for the low-energy tail of the giant dipole resonance
\cite{paa09}. 

Of course, not only the dipole, but also other multipoles could
 exhibit pronounced low-energy strength in neutron-rich nuclei. A
 theoretical study of the occurrence of pygmy quadrupole resonances
 has recently been reported in the framework the quasiparticle-phonon
 model \cite{TL.11}, and the monopole response in very exotic nuclei
 such as $^{60}$Ca was investigated using the self-consistent
 Hartree-Fock calculation plus the random phase approximation (RPA) 
 with Skyrme interactions \cite{sag97}. It was shown that near the
 drip line the monopole response could develop a very pronounced
 structure at low energy. The present study considers more realistic
 examples of nuclei in which the low-energy monopole strength could
 be measured in the near future. For the Ni isotopes, in particular,
 calculations of the monopole response in $^{78}$Ni with the Gogny-RPA
 did not predict pronounced low-lying strength \cite{per05}, as
 evidenced from the percentage of the EWSR exhausted in the low-energy
 region. In Ref. \cite{ter06} the monopole strength in Nickel isotopes
 was studied using the Skyrme-QRPA approach with the SkM* functional.
 It is, however, difficult to discern a clear pattern in the
 low-energy region because of the large artificial smoothing factor.
 Here we study in more detail the occurrence of low-lying isoscalar
 monopole strength in neutron-rich Ni isotopes. In $^{68}$Ni 
 low-lying E1 strength has recently been measured \cite{wie09}.
 A new technique has been developed
 that enables measurement of monopole strength in unstable nuclei
 \cite{mon08}. An experiment has very recently been performed at GANIL to
 measure the monopole strength in $^{68}$Ni \cite{van10}, and the
 analysis is in progress.


To minimize the model dependence of results, both Skyrme and relativistic energy density functionals are 
employed in the present analysis. Skyrme Hartree-Fock + RPA calculations are performed for the nuclei 
$^{68}$Ni and $^{78}$Ni, for which pairing effects are expected to play a negligible role in the monopole 
response. The description of Skyrme HF+RPA approach 
can be found in numerous references \cite{rin80,liu76,ber75} and will not be detailed here. We only 
mention that in this work RPA equations are solved in configuration space, and the particle-hole space 
is chosen so that the energy-weighted sum rule is fully exhausted. 
The continuous part of the single-particle spectrum is discretized 
by diagonalizing the HF Hamiltonian in a harmonic oscillator basis. An important quantity that 
characterizes a given state $\nu = (E_\nu, LJ)$ is its transition density:
\begin{equation}\label{eq:tran}
\delta\rho^\nu({\bf r}) \equiv  \langle \nu \vert \sum_i \delta 
({\bf r - r_i})\vert \tilde 0 \rangle ~,
\end{equation}
with a corresponding definition of the neutron (proton) transition density 
$\delta\rho_n^\nu$ ($\delta\rho_p^\nu$) when the summation in Eq.~(\ref{eq:tran}) is 
restricted to neutrons (protons). A smoothing factor of 600 keV is used in plots 
of strength distributions.

The monopole response of Ni isotopes is also analyzed using  
the fully self-consistent relativistic
quasiparticle random phase approximation (RQRPA) based on the
Relativistic Hartree-Bogoliubov model (RHB) \cite{Paa.03}. 
Details of the formalism can be found in Refs.~\cite{Paa.03,paa07}.  
In the RHB+RQRPA model the effective interactions are implemented in 
a fully consistent way. In the particle-hole channel effective Lagrangians
with density-dependent meson-nucleon couplings are employed~\cite{PNVR.04},
and pairing correlations are described by the pairing part of the finite-range 
Gogny interaction. Both in the $ph$ and $pp$ channels, the same interactions are used 
in the RHB equations that determine the canonical quasiparticle basis, and in the
matrix equations of the RQRPA. 

\begin{figure}[tb]
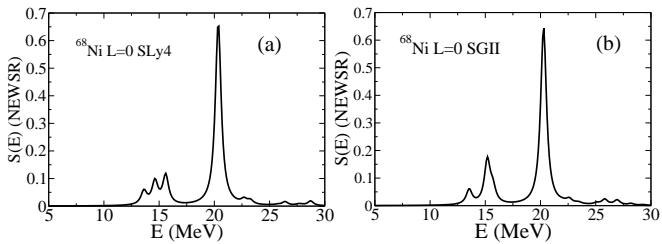

\begin{center}
\scalebox{0.17}{\includegraphics{ni68l.eps}\includegraphics{ni68g.eps}}
\caption{Skyrme-RPA isoscalar monopole strength functions in
$^{68}$Ni (in non-energy weighted sum rule units), calculated with the SLy4 (a) 
and SGII (b) energy density functionals}
\label{fig:68ni}
\end{center}
\end{figure}

\begin{figure}[tb]
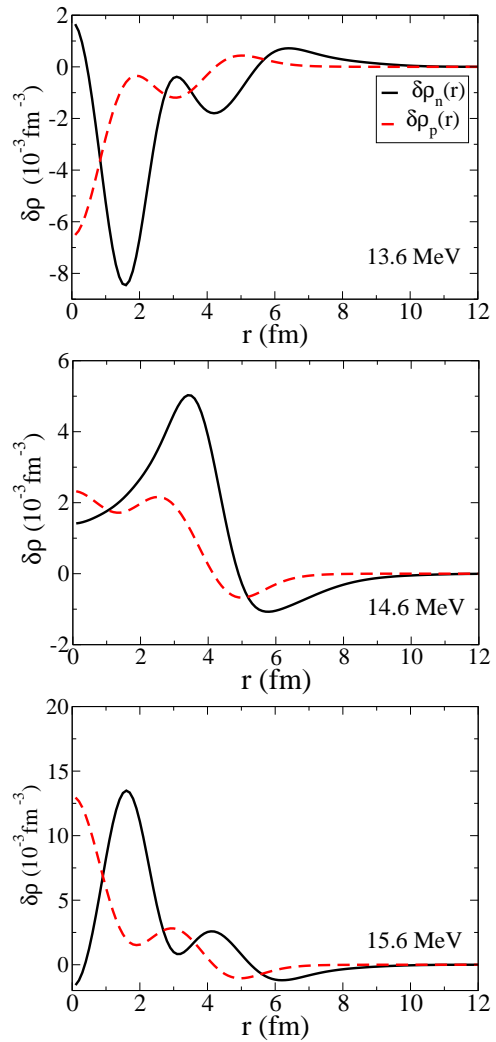

\begin{center}
\scalebox{0.25}{\includegraphics{denst.eps}}
\scalebox{0.25}{\includegraphics{denst2.eps}}
\scalebox{0.25}{\includegraphics{denst3.eps}}
\caption{(Color online) Skyrme-RPA neutron (solid line) and proton
(dashed line) transition densities for the three low-lying isoscalar
monopole peaks at 13.6 MeV, 14.6 MeV and 15.6 MeV in
$^{68}$Ni, calculated with the SLy4 functional.}
\label{fig:dens}
\end{center}
\end{figure}

In Fig. \ref{fig:68ni} we display the isoscalar monopole strength in
$^{68}$Ni, calculated with the SLy4 (left) and SGII (right)
non-relativistic functionals. The giant monopole resonance (GMR) is
calculated at about 20 MeV, and both functionals predict a pronounced
low-lying structure located between 13 and 16 MeV excitation energy.
For the response calculated with SLy4, the proton and neutron
transition densities of the three states that compose the low-energy
monopole structure are plotted in Fig. \ref{fig:dens}. These states
are not purely isoscalar: the transition densities exhibit neutron
dominated modes, and the proton and neutron densities are not in phase
in the interior of the nucleus.  The configuration analysis of these
states shows that they correspond to almost pure single hole-particle
excitations. A single configuration contributes with more than 98\% to
the total strength: neutron (2p$_{3/2}$,3p$_{3/2}$),
(2p$_{1/2}$,3p$_{1/2}$), (1f$_{5/2}$,2f$_{5/2}$) for the states
located at about 13.6 MeV, 14.6 MeV and 15.6 MeV, respectively. In nuclei 
that do not exhibit a pronounced neutron excess these
2$\hbar\omega$ unperturbed configurations are  
located at higher energies,
whereas in this case they are found below the GMR. The reason is that, because of
the neutron excess in $^{68}$Ni, the 3p2f states are located close to
the Fermi level and, moreover, the 2p1f shell is calculated just below the Fermi
level. This is not the case for the 3s2d shell. The
2$\hbar\omega$ sd configurations that build the GMR are not lowered in energy and, 
therefore, the giant resonance is located at higher energy.

\begin{figure}[tb]
\begin{center}
\scalebox{0.30}{\includegraphics{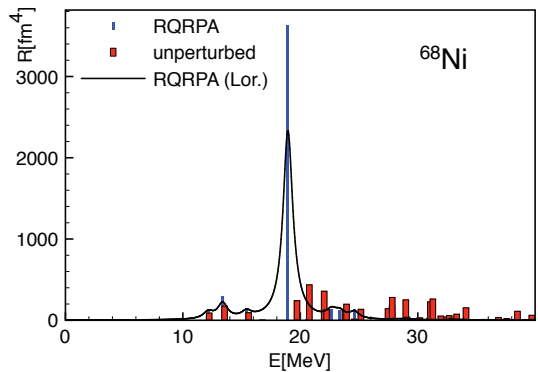}}
\caption{(Color online) Monopole response of $^{68}$Ni calculated using the RHB+RQRPA model with 
the DD-ME2 \cite{LNVR.05} functional. In addition to the RQRPA discrete spectrum and 
the corresponding Lorentzian averaged curve, the unperturbed spectrum is also displayed.}
\label{fig:ni68r}
\end{center}
\end{figure}

\begin{figure}[tb]
\begin{center}
\scalebox{0.40}{\includegraphics{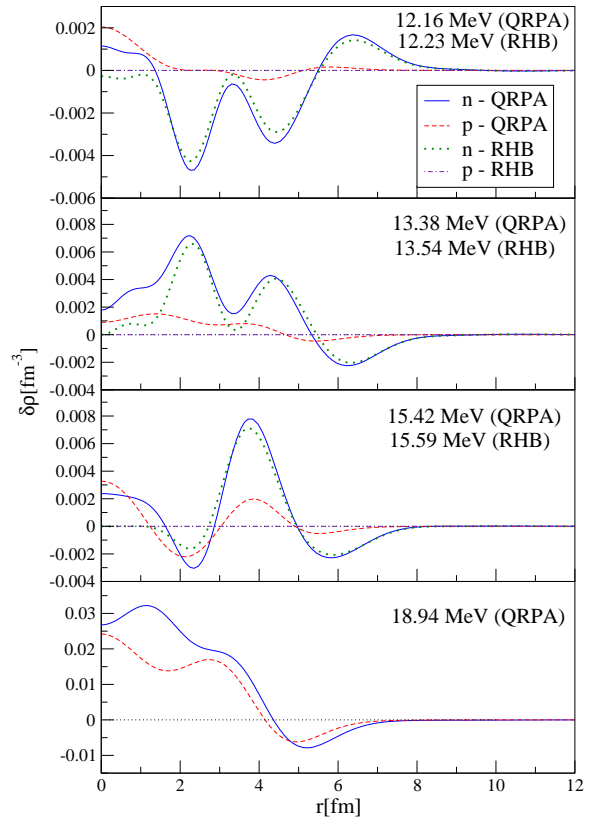}}
\caption{(Color online) $^{68}$Ni proton and neutron transition densities of the three low-lying RHB+RQRPA states 
at 12.16 MeV, 13.38 MeV, and 15.42 MeV, compared to those of the GMR at 18.94 MeV.}
\label{fig:densr}
\end{center}
\end{figure}

Fig. \ref{fig:ni68r} shows the RQPRA prediction for the isoscalar
monopole response in $^{68}$Ni, calculated with the relativistic
functional DD-ME2 \cite{LNVR.05}. Despite a small difference in the
predicted position of the GMR, the results for the low-lying part are
very similar to those obtained with the Skyrme functionals, and the
configuration analysis leads to the same conclusion about the
non-collective character of these low-lying excitations. The
comparison of the RQRPA spectrum with the unperturbed states shows
that the residual interaction affects the strength, but not the
location of the low-lying states. They are well separated from the GMR
which, in this relatively light nucleus, is found at high energy. The
corresponding proton and neutron transition densities are plotted in
Fig. \ref{fig:densr}. The transition densities for the low-lying
states at 12.16 MeV, 13.38 MeV, and 15.42 MeV (these states have the
same particle-hole structure as the corresponding ones in the
Skyrme-RPA calculation) correspond to almost pure neutron modes, and
the radial dependence is very different form the typical isoscalar GMR
transition densities exhibited by the collective state at 18.94 MeV.
The enhancement of the low-lying monopole strength in the RQRPA case
compared to the RHB one is due to a small increase of the magnitude of
the neutron transition density, the strength being a momentum of the
transition density. Also, a slight admixture of neutron and proton
configurations contribute to these states.

\begin{figure}[tb]
\begin{center}
\scalebox{0.17}{\includegraphics{ni78l.eps}
\includegraphics{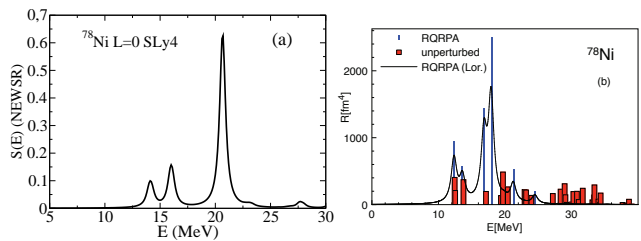}}
\caption{(Color online) Monopole response of $^{78}$Ni calculated using the Skyrme-RPA model with the 
SLy4 functional (in non-energy weighted sum rule units) (a), and the
RHB+RQRPA model with the DD-ME2 functional (b). 
 }
\label{fig:ni78}
\end{center}
\end{figure}

\begin{figure}[tb]
\begin{center}
\scalebox{0.35}{\includegraphics[angle=-90]{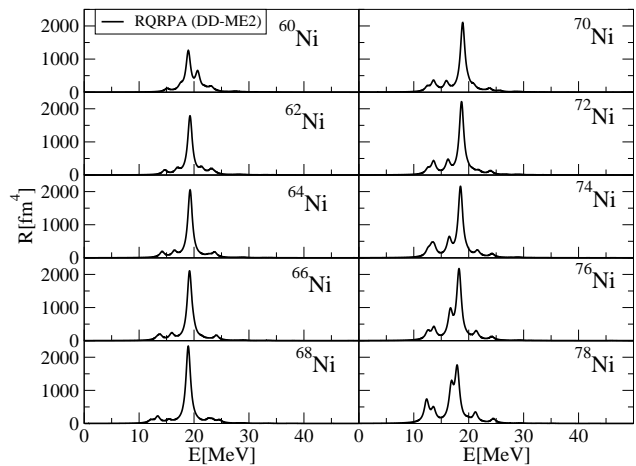}}
\caption{Monopole response of even-A $^{60-78}$Ni isotopes calculated using 
the RHB+RQRPA model with the DD-ME2 functional. }
\label{fig:ni60-78}
\end{center}
\end{figure}

Similar results are also obtained for $^{78}$Ni (Fig.
\ref{fig:ni78}), for which both Skyrme-RPA and RHB+RQRPA calculations
predict pronounced strength in the energy region around 15 MeV
excitation energy, well separated from the GMR. The low-lying strength
is more pronounced than in the case of $^{68}$Ni.  The overall
structure predicted by the DD-ME2 functional is shifted to somewhat
lower energy compared to the one calculated with SLy4. The comparison
between the unperturbed spectrum and the full RQRPA response nicely
illustrates how the residual interaction builds the collective GMR
from the states above 15 MeV excitation energy. The two low-lying
unperturbed states are not lowered in energy, even though their
strength is considerably increased by the inclusion of the residual
interaction in the full relativistic RPA. To illustrate the evolution
of low-energy monopole strength in Ni isotopes, in
Fig.~\ref{fig:ni60-78} we show the monopole response of even-A
$^{60-78}$Ni isotopes calculated with the RHB+RQRPA. One can clearly
follow how the low-lying strength in the energy region between 10 MeV
and 15 MeV develops with the neutron excess.  Since these states,
predicted both by Skyrme-RPA and RQRPA calculations, are
non-collective, their occurrence essentially depends on the position
of the neutron Fermi level in the single-neutron spectrum. The RPA
strength contained in these states, however, is markedly enhanced by
the residual interaction. These predictions point to a possibly very
interesting measurement of low-lying isoscalar monopole states in
neutron-rich Ni isotopes, that would probe almost pure single-neutron
configurations. A resolution of approximately 2 MeV can currently be achieved by
the experimental setup dedicated to the measurement of the monopole response in
unstable nuclei \cite{mon08}. An improvement to a 1 MeV energy resolution is
expected \cite{van10}, and the developpement of next generation active
targets such as ACTAR \cite{raa09} will proceed along this path. 
It should be noted, however, that the physical width of the monopole response 
(the spreading width, the decay width, and the Landau damping) could also 
prevent a clear separation between different low-energy states.

In conclusion, fully self-consistent microscopic Skyrme HF+RPA and
relativistic RHB+RQRPA calculations have been performed for the
isoscalar monopole response in neutron-rich Ni isotopes, some of 
which should become experimentally accessible in the near future. Both the
non-relativistic and relativistic models, based on Skyrme functionals
and the relativistic functional DD-ME2, respectively, predict the
occurrence of pronounced low-energy monopole states in the energy
region between 10 MeV and 15 MeV, well separated from the isoscalar
GMR. From the analysis of transition densities, transition matrix
elements and the particle-hole configurations that correspond to these
states, it is evident that the low-energy monopole states represent
almost pure neutron single hole-particle excitations. Their occurrence
can be related to the closeness of the corresponding orbitals to the
neutron Fermi level in nuclei with large neutron excess. Even though
their location is not modified with respect to the corresponding
unperturbed states, their (Q)RPA strength is considerably enhanced by
the residual interaction. The theoretical analysis predicts an
increase of low-energy monopole strength with neutron excess. A
measurement of these states would provide a direct information on the
2$\hbar\omega$ gap, and directly probe the single-particle spectrum in
exotic neutron-rich nuclei, that is known to be driven by the
spin-orbit and the tensor terms of the effective inter-nucleon
interaction. In particular, measurements of the monopole strength in
neutron-rich Nickel isotopes are currently being performed, and this
work will help to interpret the results.

\section*{Acknowledgement}
This work has been supported in part by the ANR NExEN grant, by
MZOS - project 1191005-1010, and by the Croatian Science Foundation.

\end{document}